\documentclass[smallextended]{llncs}

\usepackage{amsmath}
\usepackage{amsfonts}
\usepackage{graphicx,color}
\usepackage{subfig}
\usepackage{url}
\usepackage[linesnumbered,ruled,vlined]{algorithm2e}
\usepackage{amssymb}
\usepackage{verbatim}
\DeclareCaptionType{copyrightbox}

\newcommand{\rng}{\mathsf{RNG}}
\newcommand{\gen}{\mathsf{Gen}}
\newcommand{\setup}{\mathsf{Setup}}
\newcommand{\keygen}{\mathsf{KeyGen}}
\newcommand{\maskgen}{\mathsf{MaskGen}}
\newcommand{\computesum}{\mathsf{ComputeSum}}
\newcommand{\verify}{\mathsf{Verify}}
\newcommand{\computeunion}{\mathsf{ComputeUnion}}
\newcommand{\computeintersect}{\mathsf{ComputeIntersect}}
\newcommand{\computescalar}{\mathsf{ComputeScalar}}
\newcommand{\pk}{\textit{PK}}
\newcommand{\sk}{\textit{SK}}
\newcommand{\mak}{\textit{MaK}}
\newcommand{\aes}{\textit{aes}}
\newcommand{\ope}{\textit{ope}}

\newcommand{\grid}{\textit{gid}}
\newcommand{\rid}{\textit{rId}}
\newcommand{\bl}{\textit{bl}}

\newcommand{\enc}{\mathsf{Enc}}
\newcommand{\dec}{\mathsf{Dec}}
\newcommand{\ek}{\textit{eK}}
\newcommand{\dk}{\textit{dK}}

\newcommand{\queryCount}{\textit{QueryCount}}
\newcommand{\queryGroupBy}{\textit{QueryGroupBy}}

\begin{document}

\title{CloudMine: Multi-Party Privacy-Preserving Data Analytics Service}
\author{Dinh Tien Tuan Anh, Quach Vinh Thanh, Anwitaman Datta\\
ug93tad@gmail.com, \{vtquach,anwitaman\}@ntu.edu.sg}
\institute{School of Computer Engineering
\\
Nanyang Technological University}

\maketitle

\begin{abstract}
An increasing number of businesses are replacing their data storage and computation
infrastructure with cloud services. Likewise, there is an increased emphasis on performing
analytics based on multiple datasets obtained from different data
sources. While ensuring security of data and computation outsourced to a third party cloud is
in itself challenging, supporting analytics using data distributed across
multiple, independent clouds  is even further from trivial. In
this paper we present \emph{CloudMine}, a
cloud-based service which allows multiple data owners to perform privacy-preserved computation
over the joint data using their clouds as delegates. CloudMine protects data privacy with
respect to semi-honest data owners and semi-honest clouds. It
furthermore ensures the privacy of the computation outputs
from the curious clouds. It allows data owners to reliably
detect if their cloud delegates have been \emph{lazy} when carrying out the delegated
computation. CloudMine can run as a centralized service on a single cloud, or as a distributed
service over multiple, independent clouds. CloudMine supports a set of basic computations that
can be used to construct a variety of highly complex, distributed privacy-preserving data
analytics. We demonstrate how a simple instance of CloudMine (secure sum service) is used to
implement three classical data mining tasks (classification, association rule mining and
clustering) in a cloud environment. We experiment with a prototype of the service, the results
of which suggest its practicality for supporting privacy-preserving data analytics as a
(multi) cloud-based service. \end{abstract}


\section{Introduction}
An enormous amount of data is being generated everyday from a plethora of computing devices.
Traditionally, data is stored in the  data owner's in-house infrastructure, and access to
outsiders is provided typically through web services~\cite{xignite,noaa,resmap}.  Data from
multiple sources can be mashed-up or jointly analyzed to create new services and derive
information that cannot be realized from individual datasets~\cite{glaeser03,eubank04,abbe11}.
However, it is often desirable or even required by law to protect data privacy. Although
numerous techniques for carrying out privacy-preserving data
analytics exist (\cite{kursawe11,duan10,yang06}), we believe that for wide-scale
adoption of such techniques, it is essential to provide them as basic, out-of-the-box services
which are flexible enough, so that individual users can freely choose their respective service
providers, and yet be able to collaborate among each other.

Recent developments of cloud computing have materialized a concrete platform for rapid
realization of the service-oriented computing paradigm~\cite{wei10}. Cloud providers (Google,
Amazon, Salesforce, etc.) offer computing as a service, from which software services can be
built, sold and integrated into complex applications. Migration of private IT infrastructures
to the cloud is gathering momentum~\cite{hajjat10,tak11}, as many companies and government
agencies are moving most (or all) of their data, application logics and front-end services to
the cloud. Recent advances in cloud computing have largely succeeded in accommodating the
demand for cheap, elastic and scalable computing resources. However, security issues related to
the outsourced data and computation remain a challenging obstacle to
overcome~\cite{popa11,wang11}.

Our work is motivated by the realization of these two trends, namely the need for a service for
privacy-preserving analytics and the availability of cloud computing as a platform for
service-oriented computing. More specifically, this work concerns the design space of a
cloud-based service for carrying out distributed, privacy-preserving data analytics. We present
\emph{CloudMine}, a cloud-based, on-demand service that data owners can leverage to perform analytics
over their joint data. CloudMine runs on the cloud (or \emph{delegate}) and supports three
basic functions: secure sum, secure set union and intersection, secure scalar product.
CloudMine provides three security
assurances.  First, confidentiality of individual's data is protected from other semi-honest
data owners, as well as from colluding, semi-honest clouds. Second, outputs of the joint
computations are protected from the semi-honest clouds. Third, data owners can reliably detect
if their delegates have been lazy, i.e. if they have skipped the computations.

CloudMine can be used in a centralized manner when all the data owners use the same service on
a single cloud. More importantly, it also works well in distributed settings where
different data owners invoke different services on their delegates. In  such setting,
multiple instances of CloudMine participate in a distributed protocol in order to achieve the
same functionality. The security properties of CloudMine are still guaranteed in
this distributed environment, even when the clouds collude with each other.


A use case of CloudMine is illustrated in the following example. Suppose there is a number of
supermarkets wishing to learn customer purchase behavior by performing association rule mining
over their joint data. Each supermarket stores their customer transaction data in-house because
the data contains sensitive information, while outsourcing the rest of its IT operation to the
cloud.  Suppose the supermarkets would like to outsourcing the computation (association rule
mining) to their delegate clouds, without revealing their sensitive data to the clouds and to
each other.  Since the customer purchase behavior (output of the computation) is
valuable to the participating supermarkets, they will like it to be kept secret from the clouds
(for otherwise, the latter can benefit from the information without contributing any data).
They will also like to be able to detect if their clouds have been unscrupulous, i.e. skipping
the delegated computations while still charging them for the same. Such lazy behavior could
undermine accuracy of the final result. CloudMine meets these functionality and security
requirements, and it can be readily invoked on the clouds. First,
CloudMine supports set intersection and sum operations, which can be used to carry out
association rule mining~\cite{clifton02}. Next, CloudMine protects confidentiality of data
owner's input from other data owners and from the clouds, thus the supermarkets can be assured
of the data privacy from each other and from the clouds.  In addition, CloudMine
protects output of the computation from the clouds, therefore the result from association rule
mining is only learned by the participating supermarkets.  Finally, CloudMine allows data
owners to detect lazy clouds, thus the supermarkets can use CloudMine to verify if their clouds
have been unscrupulous.

Privacy-preserving data analytics is an active area of research. Existing techniques are based
on either a generic secure multi-party computation~\cite{yao82,yang06}, or on using a
semi-honest third party~\cite{duan10,kursawe11}. Our work distinguishes itself from the former
in that the clouds are used as delegated computation units, hence it is more scalable. It
differs from the latter in that we consider a stronger adversary model for the cloud delegates.
Especially, we consider colluding adversaries who try to learn both the inputs and outputs of
the computation while doing as little as possible. Furthermore, while previous works consider
ad-hoc sets of data analytic tasks, each focusing on one primitive function (mostly secure sum
function), CloudMine is designed as a service with a large set of analytic functions including
secure sum, set operations and scalar products. We defer more detailed discussion to the next
section.

The key enabling technique used in CloudMine is additive homomorphic
encryption~\cite{paillier99}, which allows data owners to encrypt their private inputs before
exporting it to the CloudMine cloud delegate. The ciphertexts contain additional information to
allow for verification of computation. Secure set operations (intersection and union) are
reduced to secure sum operations by encoding set membership into the plain-text inputs. Scalar
product is computed by leveraging the homomorphic property of Paillier encryption and the
secure sum function. In all cases, the keys are kept secret from the clouds, hence
they are unable to decrypt the outputs.




Our contributions are as follows:
\begin{enumerate}
\item We present a model for cloud-based services for distributed, privacy-preserving data
analytics. The model allows data owners to outsource their private computations to the cloud in
a privacy-preserved manner.

\item We describe how the service can be implemented to support a number of cardinal data analytic functions, namely secure sum, secure set operations, and secure scalar product. We name the service
CloudMine.

\item We demonstrate how CloudMine can be used for more complex data mining tasks --- namely
classification, association rule mining and clustering --- in a hybrid cloud setting. In
particular, we show how CloudMine works when some parts of the data are stored in encrypted
form in the public clouds.

\item We benchmark CloudMine on a cloud platform, both as a stand-alone service and as
a part of more complex data mining applications. The results suggest that the overheads incurred
because of the added security mechanism are reasonable and amortized as the workload increases.
They indicate that it is practical to outsource distributed, privacy-preserving data
analytics to a (multi) cloud service.
\end{enumerate}

In the next section, we discuss in detail the system and adversary model of
CloudMine. Section~\ref{sec:protocol} delineate the CloudMine protocols for various
analytic functions. Section~\ref{sec:dataMining} describes how three classic data mining tasks
can be built using an instance of CloudMine in a hybrid cloud setting. Section~\ref{sec:evaluation} follows
with experimental evaluation before related works are discussed in
Section~\ref{sec:relatedWork}. We conclude and outline some planned future work in Section~\ref{sec:conclusion}.

\section{CloudMine Model}
\label{sec:model}
\subsection{System Model}
The system using CloudMine consists of two kinds of entities: \textbf{\emph{data owners}} (or
\textbf{\emph{parties}}) and
\textbf{\emph{clouds}} (or \textbf{\emph{delegates}}). The data owners $\mathbb{P} =
\{P_0,P_1,..,P_{n-1}\}$ wish to compute a
function $f(x_0,x_1,..,x_{n-1})$ where $x_i$ is the input of $P_i$, without
revealing the input to each other. The clouds $\mathbb{C} = \{C_0,C_1,..,C_{k-1}\}$ where $k\leq
n$ are the service providers. Each party uses one of these cloud, and each cloud is utilized by
at least one party. Denote $\delta(i) \in \mathbb{C}$ as the delegate used by party $P_i$. 

At a high level, data owners use CloudMine in two steps in order to compute $f(.)$. First, they
enter the \emph{setup} phase, in which they agree on a function $\phi$ (and $\phi^{-1}$) and a
secret $\textit{sk}$.  Next, each party $P_i$ computes $\phi_\textit{sk}(x_i)$ and
sends it to the delegate $\delta(i)$. In turn, the delegates exchange messages among themselves and
effectively compute $\pi = \phi_\textit{sk}(f(x_0,x_1,..))$. The data owners receive $\pi$ from their
respective delegates and compute $\phi_\textit{sk}^{-1}(\pi)=f(x_0,x_1,..)$.

\subsection{Adversary model}
\textbf{Data owners / parties}  are \emph{curious but honest}. They follow the protocol for computing
$f(.)$ correctly, but passively try to learn the private inputs of each other. They could collude with
each other, but the number of colluding parties is less than $n-1$.

\textbf{Clouds/Delegates} are \emph{curious and lazy}. They are curious with respect to the
parties' private inputs as well as the output of $f(.)$. They do not actively subvert the
computation, but are lazy in the sense that they try to do as little as possible while
charging the data owners for the same. For example, they may skip some (or all) of the
computations, replay results from the previous rounds, or even replace inputs from the data
owners with other values in order to avoid computation. This model is justified by the
economic incentives of the cloud providers to over-charge customers without being
detected~\cite{wang11a}, as well as the legal realities in which the clouds can sniff sensitive
information without the liability of committing a criminal offense.

The collusion between parties and delegates is weak. In particular, the delegates may
reveal the messages exchanged during the computation of $\phi(f(.))$ to the
parties, but the shared secret between the parties are not revealed to the delegates.
If the shared secret is revealed, it is not possible to guarantee
privacy of the computation output. 

\subsection{Security goals}
Given the model above, CloudMine aims to provide the following security
assurances:
\begin{enumerate}
\item Data owners cannot learn each other's private inputs.
\item Delegates cannot learn the parties' private inputs, nor can they learn the output $f(.)$.
\item Delegates cannot skip, replay or replace inputs of the delegated computations without
being detected by the parties.
\end{enumerate}

\subsection{Discussion}
Existing works on multi-party private computation, which underlie privacy-preserving data
analytics, can be grouped into two different approaches. The first is based on secure
multi-party computation, in which data owners interact with each other directly to evaluate
a function based on their private inputs. For example, \cite{vaidya03,yang06} use generic
multi-party computation circuits~\cite{yao82}. The second approach is
based on a third party, in which data owners send their encrypted inputs to the third party
which evaluates the function. \cite{duan10,shi11,kursawe11}, for instance, follow this approach.

Our model differs to the secure multi-party computation approach mainly in that the parties 
delegate their computations to the clouds. As a result, instead of interacting with each other,
which does not scale well with the size of $n$, each party only interacts with its delegate.
More importantly, this model allows for much more efficient implementation of the private
computation than using generic, circuit evaluation (which
takes in the order of seconds to compute a 2-party secure sum~\cite{duan10}).

Our model share some similarities with the second approach. On one hand, when
$k=1$, the system model of CloudMine is the same as in many other works which rely on a single
third party. On the other hand, CloudMine distinguishes itself in a number of aspects. First,
we consider the case when there are multiple, independent third parties that each data owner
can individually choose to use as delegate. CloudMine is designed to
resist collusion among these delegates. This is different from~\cite{duan10} which also
supports multiple servers, but they are assumed to be non-colluding. Second, CloudMine
adversary model considers the delegates trying passively to learn the output of $f(.)$,
which is not the case in previous work. We believe such outputs
may leak sensitive information. For example, the clouds may use the aggregate (sum) values
together with off-line knowledge to derive sensitive information~\cite{dwork06}, or they may
directly infer parts of the data owners' private inputs from the output of the set intersection
function.  Third, we consider the clouds to be lazy which may skip the delegated computation
and subsequently render the output $f(.)$ incorrect. This behavior presents a realistic threat
to the utility and integrity of the analytics, yet it has not been addressed in existing works.

Finally, designing CloudMine as a service on the cloud has another benefit with respect to
scalability. Since the cloud maintains the service, it can monitor the workload and
automatically add more resources to deal with increases in workload. This automatic, seamless
scaling is an essential practical improvement over systems such
as~\cite{duan10} which require complete reconfiguration and re-run of the protocols to
accommodate more servers.

\section{CloudMine Implementation}
\label{sec:protocol}
We now describe how to implement the CloudMine service to support three analytic functions:
secure sum, secure set operations (intersection and union), and secure scalar product. These
primitives serve as a powerful toolbox for doing privacy-preserving analytics, ranging from
database queries such as join~\cite{narayan12} and aggregate~\cite{kursawe11} to complex mining algorithms
such as collaborative filtering~\cite{duan10}.

CloudMine relies on an \emph{additively homomorphic} encryption scheme to protect privacy of
the data owners' inputs and to implement the basic secure sum function. In particular, we use
Paillier~\cite{paillier99}, a randomized encryption scheme consisting of three algorithms
$(\mathsf{Gen}, \enc, \dec)$ where $\enc$ and $\dec$ are encryption and decryption algorithms
which use the key generated by $\mathsf{Gen}$. Paillier has the following property: 
\[
\enc(\ek,m_1).\enc(\ek,m_2) =\enc(\ek,m_1+m_2)
\]
where $\ek$ is the encryption key. Compared to other additively homomorphic schemes based on
Elgamal~\cite{ugus09}, Paillier requires longer bit-length. But we can overcome this by packing
multiple inputs into a single plaintext so that they can be encrypted and decrypted at the same
time~\cite{popa11}. Suppose the inputs are at most $b$ bits and Paillier's plaintexts are $\bl$
bits. Suppose further that any sum value is smaller than $2^{t+b}$ for some values of $t$, then we can pack $c$ inputs
into a single plaintext (where $c \leq \lceil \frac{\bl}{b+t} \rceil$) as follows:
\[\langle x_1 \| x_2 \|..\| x_c \rangle = z \| x_1 \| z' \| x_2 .. \| z' \| x_c
\]
where $z'$ contains $t$ bits of $0$ and $z$ contains $(\bl-c.(t+b))$ bits of $0$.

In the following, we describe the construction of three services that constitute CloudMine. The
\emph{secure sum service} implements the aggregate function, \emph{secure set service} the set union and
intersection function, and \emph{secure scalar service} the scalar product function. 

\subsection{Secure Sum Service}
\label{subsec:securesum}
The secure sum service, denoted as $\mathsf{S}_\textit{sum}$, consists of five 
protocols: $\mathsf{S}_\textit{sum} = (\setup,\keygen,\maskgen,\computesum,\verify)$. The first
three protocols are performed once at the beginning, while $\computesum$ and $\verify$ are
invoked for each round of computation.  

\begin{itemize}
\setlength{\itemsep}{0.3cm}
\item $\setup(\kappa)$: generate public parameters with $\kappa$ being
the security parameter. The result is the tuple:
\[
\textit{PK} = (\grid,\rng,\mathbb{G}_p(g), \gen,b,\bl)
\]
where $\grid$ identifies the group to which all parties belong, $\rng$ is a random number
generator, $\mathbb{G}_p(g)$ is an algebraic group of prime order $p$ and generator $g$, $\gen$ is an
algorithm for generating Paillier keys. $b$ and $\bl$ are
bit lengths of the inputs and Paillier plaintexts respectively. 

\item $\keygen(\textit{PK})$: data owners execute this protocol to establish a
share secret:
\[
\textit{SK} = (\ek,\dk,\rid)
\] 
where $(\ek,\dk)$ is a Paillier key pair and $\rid$ is a random number identifying the initial round
of computation. First, the parties follow the protocols as proposed in~\cite{burmester05} to
generate a secret $x \in \mathbb{G}_p$ using the clouds and without the latter learning $x$. Next,
they use $x$ as the seed to initialize the random number
generator $\rng$, which is then used by $\gen$ to generate $(\ek,\dk)$. Finally, the parties assign
the next random number generated by $\rng$ as $\rid$.

\item $\maskgen(i, \textit{PK})$: each party $P_i$ invokes this protocol to generate its own
secret 
\[
\textit{MaK}_i = (r_i, \eta_i)
\]
such that $r_i, \eta_i$ are random values from an algebraic group of specific size, $\eta_i \neq 0$
and the sum of $r_i$ and $\eta_i$ across all data owners are known, i.e. $\sum_i r_i = \sum_i \eta_i
= 0$. 

First, $P_i$ creates random values $r_{ij}$ and $\eta_{ij}$ (using its private source of
randomness) for all $P_j$ ($i \neq j$) belonging to the group $\grid$.  Next,
$r_{ij}$ and $\eta_{ij}$ are encrypted with $P_j$'s public key and sent to $\delta(i)$ which
subsequently forwards them to $P_j$.  Having received the encrypted $r_{ji}, \eta_{ji}$ from its
delegate, $P_i$ then computes 
\[
r_i = \sum_{i \neq j}(r_{ij} - r_{ji}) \qquad \eta_i = \sum_{i \neq j}(\eta_{ij} -\eta_{ji})
\]
It can be seen that $P_j$ cannot learn $r_i, \eta_i$ for $i \neq j$, and that
$\sum_i r_i = \sum_i \eta_i = 0$.

\item $\computesum(i,\textit{PK}, \textit{SK}, \textit{MaK}_i, x_i)$: each party $P_i$
constructs the ciphertext $c_i$ for its private input $x_i$ as follows:
\[
c_i = \enc(\ek, \langle\eta_i||\rid||(x_i+r_i)\rangle)
\]
It then sends $c_i$ to its delegate which then broadcasts it to the other delegates. Finally, each
delegate computes:
\[
c = \prod c_i = \enc(\ek, \langle\sum \eta_i || n.\rid || \sum x_i\rangle)
\]
and forwards it to the party. Finally, $P_i$ invokes $\verify(i,\textit{PK}, \textit{SK},
\textit{MaK}_i, c)$. If the result of this verification protocol is $y \neq \bot$, the party
returns $y$ as the final sum.   

\item $\computesum(i,\textit{PK}, \textit{SK}, \textit{MaK}_i,\overline{x_i})$: takes as parameter a vector of
inputs $\overline{x_i}$ instead of a single input. Let $s = \lceil \frac{\bl-2(b_r+\lceil
log_2 n \rceil)}{b_r+b}\rceil$ where $b_r$ is the bit length of $\eta_i$ and $\rid$. For $0 \leq k <
\lceil\frac{b.|\overline{x_i}|}{s}\rceil$, we construct message $m_k$ as follows:
\[m_k = \langle x_{k.s}\|x_{k.s+1}\|..\|x_{(k+1).s-1} \rangle\]
where $|m_k| = \bl-2(b_r+\lceil log_2 n \rceil)$ bit. For
each $m_k$, the party invokes $\computesum(i,\textit{PK}, \textit{SK}, \textit{MaK}, m_k)$. If
the result $y \neq \bot$, it extracts the sums $y_{k.s}, y_{k.s+1},..,y_{(k+1).s-1}$ from $y$. After each
invocation, the party increments $\rid$ and updates $\sk$ accordingly.  

\item $\verify(i, \textit{PK}, \textit{SK}, \textit{MaK}_i, c)$: each party decrypts the ciphertext
$c$ and checks that the result is of the following form: 
\[
\dec(\dk,c) = \langle 0 || n.\rid || y \rangle
\]
If true, $y$ is returned as the final sum. 
\end{itemize}

\subsubsection{Discussion.}
We now discuss how the protocols above meet the security requirements  
listed in Section~\ref{sec:model}. First, data owners cannot learn each other's inputs, because
each input $x_i$ has been masked with a secret value $r_i$. Second, delegates cannot
extract the sum $\sum_i x_i$ from the ciphertext $c$, because they do not have access to the
decryption key $\dk$. Third, delegates cannot replay old values without being detected, since
each ciphertext is embedded with a fresh value of $\rid$. They cannot replace $c_i$ with another
valid ciphertext either, because they do not have access to $(\ek, r_i, \eta_i)$, thus invalid ciphertexts
will be detected by the  verification protocol. Neither can they skip some (or all) of the
inputs for the computation of $c = \prod c_i$, because it will cause verification to fail,
since $\dec(c) \neq \langle 0||n.rId||y \rangle$. Finally, each delegate can compute $c = c_i^{n}$ (raising to the power of
$n$ may be cheaper than $n$ multiplications), which makes the second element of $\dec(c)$
to be the same as $n.\rid$.  However, verification will still fail, because $n.\eta_i \neq 0$.

Security of $\mathcal{S}_\textit{sum}$ depends on the fact that delegates do not know the
shared secret $\textit{SK}$ or the data owner secret $\textit{MaK}_i$. Every party $i$ must protect
 $\textit{MaK}_i$ from other parties . To ensure long-term security, it is important to refresh
$\textit{SK}$ as well as $\textit{MaK}_i$, albeit refreshing the latter can be done after
longer intervals. This can be achieved by invoking $\keygen$ and
$\maskgen$ again. Alternatively, if $P_i$ stores the original $\{r_{ij}, r_{ji}, \eta_{ij},
\eta_{ji}\,|\, j \neq i\}$, the new $\textit{MaK}_i$ can be computed as: 
\[r_i' = \sum_{i \neq j}(H(r_{ij}) - H(r_{ji})) \qquad \eta_i' = \sum_{i \neq j}(H(\eta_{ij}) - H(\eta_{ji}))
\]
where $H$ is a cryptographic hash function. 

The verification of delegate behavior relies on the party encoding its secret $\mak_i$ to the
ciphertexts. As a consequence, the memory overhead is $ o = \frac{2.(b_r+log_2n)}{\bl}$, which
decreases as the Paillier bit-length $\bl$ increases. The $\verify$ protocol is performed at
the end of every $\computesum$ protocol. This can become overhead when there are many rounds of
computations. Hence, we extend $\mathcal{S}_\textit{sum}$ to allow parties to invoke $\verify$
only with a probability $p$.  The probability of successfully detecting consistent misbehavior,
$p_v$, can be made arbitrarily high after a number of verification. Specifically, $p_v = 1-(1-p)^{n.k}$ where $k$
is the number of random checks.

\subsection{Secure Set Service}
The service for secure set union and intersection can be built directly from the secure sum
service. Intuitively, the input sets are encoded into plaintext messages which are used as
inputs for $\mathcal{S}_\textit{sum}$. The union or intersection set is then decoded from the
final sum values.

The secure set service, denoted as $\mathcal{S}_\textit{set}$, consists of five protocols: 
\\
$\mathcal{S}_\textit{set} =
(\setup,\keygen,\maskgen,\computeunion,$\\ $\computeintersect)$. 
\begin{itemize}
\setlength{\itemsep}{0.3cm}
\item $\setup$, $\keygen$, $\maskgen$ are the same as in the secure sum service,
except that the public parameter $\textit{PK}$ also contains a universal domain $U$.  
 
\item $\computeunion(i,\textit{PK}, \textit{SK}, \textit{MaK}_i, \overline{x_i})$: each party
inputs a vector $\overline{x_i} \in U^*$ and computes the union set as follows.  A vector $I =
(a_0, .., a_{|U|-1})$ is constructed, in which $a_i = 1$ if $U[i] \in \overline{x_i}$ and $a_i=0$
otherwise. The party then invokes $\computesum(i,\textit{PK}, \textit{SK}, \textit{MaK}_i, I)$. When
the secure sum service returns $s_0,s_1,..,s_{|U|-1}$, it computes $\{U[i] \,|\, s_i
\geq 1\}$ as the union set.  

\item $\computeintersect(i,\textit{PK}, \textit{SK}, \textit{MaK}_i,\overline{x_i})$ works in the
same way as $\computeunion$, except that the intersection set is computed as $\{U[i] \,|\, s_i = n\}$.
\end{itemize}

\subsubsection{Discussion.} 
This service has the same security properties as for secure sum. The number of encryptions per
set operation is $\lceil \frac{|U|(log_2n+1)}{\bl-2(log_2n+b)} \rceil$, which grows linearly
with the size of $U$.  Consequently, our protocols may not scale well when $U$ and $n$ are very
large (for example, in orders of millions as in the case of large-scale collaborative
filtering). Other protocols for private set operations which scale more gracefully
(\cite{huang12,narayan12}) do not apply to our delegate model. In practice, many applications
involving secure set operations have small- to medium-size $U$ (in orders of ten or
hundred)~\cite{uciDataset}, which renders our protocols practical. For example, for $100$ data
owners, $|U|=1000$, $\bl=1024$, $b=16$, the protocols need only $9$ encryptions. We believe
that for current applications, this cost is reasonable.

\subsection{Secure Scalar Service}
Data owners are divided into two disjoint groups $X$, $Y$. Let $\overline{x}$ and
$\overline{y}$ be two vectors in which $x_i$ is the private input of party $X_i$, $y_i$ the
private input of $Y_i$. The secure scalar service, $\mathcal{S}_\textit{sp}$ allow the data owners in both groups to compute
\[
p = \overline{x}.\overline{y} = x_0.y_0+x_1.y_1+..+x_\frac{n-1}{2}.y_\frac{n-1}{2}
\]

$\mathcal{S}_\textit{sp}$ consists of four protocols $(\setup, \keygen, \maskgen,
\computescalar)$, and it also makes use the secure sum service.
\begin{itemize}
\setlength{\itemsep}{0.3cm}

\item $\setup(\kappa)$ is the similar to that in $\mathcal{S}_\textit{sum}$. It outputs public
the parameter:
\[
\pk = (\grid_x, \grid_y, \rng, \mathbb{G}_p(g), \gen, b, \bl)
\] 
where $\grid_x$ and $\grid_y$ are identities of group $X$ and $Y$ respectively. 

\item $\keygen(\grid, \pk)$: parties that belong to the group $\grid$ execute this protocol to
establish  a shared secret among them. The protocol is the same as in
$\mathcal{S}_\textit{sum}$, and the result is  
\[
\sk_\grid = (\ek_\grid,\dk_\grid,\rid_\grid)
\] 

\item $\maskgen(i,\grid,\pk)$: each party $i$ in group $\grid$ first generates a
Paillier key pair $(\ek_{\grid,i}',\dk_{\grid,i}')$. Next, it generates two values $r_{\grid,i},
\eta_{\grid,i}$ in the same way as in $\mathcal{S}_\textit{sum}$, i.e. $\eta_{\grid,i} \neq 0$
and $\sum_i \eta_{\grid,i} = \sum_i r_{\grid,i} =0$. Denote 
\[
\mak_{\grid,i} = (\dk_{\grid,i}, r_{\grid,i},\eta_{\grid,i})
\]
as the secret of party $i$ in group $\grid$. Also, let
\[
\pk' = \{\ek_{\grid_x,i}'\} \ \cup \ \{\ek_{\grid_y,i}'\}
\]
be another set of public parameters. 
 
\item $\computescalar(i,\grid,\pk,\pk',\sk_\grid,\mak_{\grid,i},x_i)$: party $i$ in group $\grid$ executes this protocol to compute the global scalar product. Suppose
the party is $X_i$ (belonging to group $X$, and $\grid = \grid_x$), the protocol proceeds as
follows: 
\begin{enumerate}
\item $X_i$ sends $m_i = \enc(\ek_{\grid_y,i}',x_i)$  to its delegate which then forwards it to $Y_i$.
\item $Y_i$ computes $c_i = m_i^{y_i}.\enc(\ek_{\grid_x,i}',r_{\grid_y,i})$ and sends it back to $X_i$ via its delegate.
\item $X_i$ computes $z = \dec(\dk_{\grid_x,i}',c_i) = x_i.y_i + r_{\grid_y,i}$. It then invokes
the service sum service $\computesum(i,\pk,\sk_\grid,\mak_{grid,i},z)$, the result of which is the scalar product.
\end{enumerate}
\end{itemize}

\subsubsection{Discussion.}
The intuition behind $\computescalar$ protocol is for $X_i$ to compute the value
$(x_i.y_i+r_{\grid,i})$ without knowing $y_i$. This is then aggregated with the other
values from $X_j$ $(i \neq j)$ to cancel out $r_{\grid_y,i}$ and obtain the scalar product. This works
because of the homomorphic property of
Paillier and the fact that $\sum r_{\grid_y,i} = 0$.

The party $X_i$ cannot learn input of $Y_i$, because the sum $x_i.y_i$ is
masked by a random value $r_{\grid,i}$. Neither can $X_i$ learn the input of $X_j$ ($i \neq j$)
due to the property of the secure sum service.  The delegates can neither learn the intermediate
sum $x_i.y_i$ because they are encrypted with data owners' keys, nor the final scalar product
because $\mathcal{S}_\textit{sum}$ does not reveal the final sum to the delegates. In $\computescalar$, the delegates play two roles: forwarding messages between
parties and performing secure sum computations. For the former, the delegates cannot be lazy without being detected,
because messages are acknowledged (so they cannot be skipped) and freshly signed (so they cannot be
replayed). For the latter, the secure sum protocol ensures that lazy behavior will be reliably
detected.

It can be seen that delegates are more involved in this service than in
$\mathcal{S}_\textit{sum}$ or in $\mathcal{S}_\textit{set}$. In particular, they must keep
track of the group to which each party belongs, and must forward messages to the correct
delegates. This management task, if left to the data owners, may become impractical for large
systems. Since CloudMine is a cloud-based service, such tasks can be performed by the cloud in
a scalable way.

\section{Data Mining in Hybrid Clouds}
\label{sec:dataMining}
The hybrid cloud model, in which the user utilizes the combined resources of its private
infrastructure (or private cloud) and a public cloud, helps ease the transition from in-house to public-cloud computing. This
model is motivated by the need to optimize cost and performance, to cater for different demand
patterns, or to mitigate risks~\cite{eucalyptus}. In this section, we demonstrate how
CloudMine's secure sum service can be used to implement distributed, privacy-preserving data
mining algorithms in this hybrid environment.

We consider data owners as hybrid-cloud users, who partition their data into two parts: the
sensitive part maintained in the private cloud, and the less sensitive part stored in a public
cloud~\cite{zhang11}. For example, data generated by an intrusion detection system may consist
of highly sensitive records associated with the internal system, whereas traffic to/from the
front-end servers may be regarded as less sensitive. Another example is in large scale genomic
sequencing: an individual's DNA sequence is highly sensitive and must be handled in the private
cloud, whereas a reference genome can be considered as less sensitive and therefore can be
encrypted and outsourced to a public cloud~\cite{chen12}. Note that \emph{less sensitive} is
not the same as \emph{non-sensitive}, in the sense that data owners still want to have some
levels of privacy with the less sensitive data. We distinguish two \emph{logically} separate
delegates: a \emph{computation delegate} which runs CloudMine service, and a \emph{data
delegate} which maintains the owner's data. They may belong to the same cloud, or each to a
different cloud.  Adversary model for the computation delegates is the same as in the previous
section. Adversary model for the data delegates adversary model is also curious-and-lazy. In
particular, they try to learn the data stored on the public clouds, and try to do as little as
possible when answering data queries from the owners. They may collude with each other, but
they will not tamper with the data.

To protect the outsourced data from curious delegates, an encryption scheme must be used. In
our design, we employ two encryptions scheme: AES and Order-Preserving Encryption
(OPE)~\cite{boldyreva09,popa13}. AES is a deterministic scheme that supports equality comparison of
ciphertexts. OPE offers weaker security guarantees, but it supports inequality comparison of
ciphertexts, which can be used for range queries. For the sake of simplicity, we store two
encrypted copies of the data on the data delegates (a more elegant approach can be found in
CryptDB~\cite{popa11}). We use the OPE scheme from~\cite{boldyreva09}, which is a stateless
encryption and does not require a third-party server (as in~\cite{popa13}). 

Untrusted data delegates necessitate protocols for ensuring query assurance. In the literature,
techniques for query assurance are probablistic which make use of redundant query execution
(\emph{ringer schemes})~\cite{sion05,du04,le12}. In this work, we use a mechanism based
on~\cite{sion05}, in which the data owner maintains a random, small portion of the outsourced data
in its private cloud. Queries to the delegates are extended with a number of fake queries, and the
results are probabilistically checked by querying the local copy of the data. Our experiments show
that maintaining as little as $15-20\%$ of the outsourced data locally is sufficiently effective to
detect lazy delegates after a small number of checks.

In the following, the data mining algorithms are run on the private cloud of each data owner.  We
assume, for simplicity, that data is in relational format and every attribute belongs to a
non-negative integer domain.  The algorithms consist of an iterative process of querying the
public-cloud database, combining it with the local data, and using the result as inputs to the
secure sum service. The fact that outputs from the interactions with the data delegates are used
during the computations involving cloud delegates may appear to be a risk to privacy, especially
when data and computation delegates collude (which is immediate when they belong to the same cloud
provider).However, privacy is ensured for two reasons. First, the computation delegates cannot learn
the data owners' inputs to the private computation, because the inputs are obtained over both the
data stored in the private cloud and data outsourced to the data delegate. Hence, results from
querying the data delegates only contribute partly to the inputs. Second, and more importantly, even
if all the data is outsourced, the delegates cannot collude and compute analytics by themselves,
because both the data and the meta-data (column names, table names, etc.) are encrypted.

\subsection{Classification (Naive Bayes).}
\begin{algorithm}
\footnotesize
\caption{Naive Bayes classification}
\label{alg:naiveBayes}
\textbf{Input}: $Y,A,V,i$\\
\textbf{Output}: $N,\{N_y\},\{N_{y,a,v}\}$\\

\vspace{0.1cm}
$\pk \leftarrow \setup(\kappa)$; $\sk \leftarrow \keygen(\pk)$; $\mak_i \leftarrow
\maskgen(i,\pk)$\\

\vspace{0.1cm}
\textbf{foreach} $y \in Y, a \in A, v \in V_a$: \\
$\quad$ $N_y^i \leftarrow \queryCount(\textit{label} = y)$ \\
$\quad$ $N_{y,a,v}^i \leftarrow \queryCount(a = v , \textit{label} = y)$ \\

\vspace{0.1cm}
\textbf{foreach} $y \in Y, a \in A, v \in V_a$:\\
$\quad$	$N_y \leftarrow \computesum(i,\pk,\sk,\mak_i,N_y^i)$\\
$\quad$ $N_{y,a,v} \leftarrow \computesum(i,\pk,\sk,\mak_i,N_{y,a,v}^i)$\\

\end{algorithm}

A classification algorithm takes as input a set of labeled, training data and outputs a
\emph{classifier} that can be used to assign label to new data. Let $N$ be the number of data
instances, $Y$ the set of labels, $A$ the set of attributes and $V_a$ the attribute domain for
$a \in A$. The NaiveBayes algorithm shown in Algorithm~\ref{alg:naiveBayes} computes:
\[
\textit{classifier} = (N, \{N_y \,|\, y \in Y\}, \{N_{y,a,v} \,|\, y \in Y, a \in A, v \in V_a\})
\] 
The label for a new instance $x$ is:
\[
\textit{label}(x) = \textit{argmax}_y(\frac{N_y}{N}.\prod_i
\frac{N_{y,i,x_i}}{N_y})
\] 

The protocol $\queryCount(a_1=v_1, a_2=v_2..)$ encrypts $a_1$, $v_1$ with AES and issues a SQL query
of the form 
{\small
\begin{align*}
&\texttt{select COUNT from } \enc_\aes(\texttt{Data})\\
& \qquad \qquad \texttt{ where } \enc_\aes(a_1) =\enc_\aes(v_1)\\
& \qquad \qquad \quad \ \texttt{ AND } \enc_\aes(a_2) = \enc_\aes(v_2) \  ..
\end{align*}
}
to the data delegate. The delegate executes the SQL query over the encrypted
data and returns the result which is probabilistically verified by the owner.

\subsection{Clustering (K-Mode).}
\begin{algorithm}
\footnotesize
\caption{K-Mode Clustering}
\label{alg:kmeans}
\textbf{Input}: $k,A,i$\\
\textbf{Output}: $M = \{m_1,..,m_k\}$

\vspace{0.1cm}
$\pk \leftarrow \setup(\kappa)$; $\sk \leftarrow \keygen(\pk)$; $\mak_i \leftarrow
\maskgen(i,\pk)$\\
Initialize $m_j = \{j,j,..,j\}$ for $m_j \in M$
$C^i = \emptyset$\\

\vspace{0.1cm}
\textbf{foreach} $m_j \in M$:\\
$\quad$ $C_{m_j}^p = \emptyset$\\
$\quad$ \textbf{foreach} $a \in A$\\
$\quad\quad$ $C_{m_j}^i(a) \leftarrow \queryGroupBy(a,m_j,M)$\\
$\quad\quad$ $C_{m_j}^i = C_{m_j}^i \, \cup\, C_{m_j}^i(a)$\\
$\quad$ $C^i = C^p \,\cup\, C_{m_j}^i$\\

\vspace{0.1cm}
\textbf{foreach} $C^i[j] \in C^i$:\\
$\quad$ $C[j] \leftarrow \computesum(i,\pk, \sk, \mak_i, C^i[j])$\\

$C^i \leftarrow C$\\
\textbf{foreach} $m_j \in M, a \in A$:\\
$\quad$ $m_j(a) \leftarrow \textit{Mode}(C_{m_j}^i(a))$\\

\vspace{0.1cm}
Repeat Step 5 until $M$ converges.
\end{algorithm}
A clustering algorithm partitions the data into separate \emph{clusters} such that distance between members of the same
cluster is smaller than that between members of different clusters. The K-Mode algorithm
(Algorithm~\ref{alg:kmeans}) finds $k$ clusters identified by their centroids (or \emph{modes})
that minimizes the dissimilarity between members of the same cluster (the $\textit{Mode}$ function).
The algorithm works in multiple rounds until the set of modes converges.

We use Manhattan distance to quantify the distance from a data instance $x$ to a mode
$c$, i.e. $\Delta(x, c) = \sum_i |x_i - c_i|$. The protocol $\queryGroupBy(a,m_i,M)$ queries
the data delegate for a list of frequencies for attribute $a$ in the portion of data
closest to the centroid $m_i \in M$. The query has the form:
{\small
\begin{align*}
& \texttt{select } \enc_\aes(a), \texttt{ COUNT from } \enc_\aes(\texttt{Data}) \texttt{ as freq}\\
& \text{\texttt{where }} \Delta(\enc_\ope(a),\enc_\ope(m_i)) < \Delta(\enc_\ope(a),\enc_\ope(m_0))\\
& \quad \texttt{ AND } \Delta(\enc_\ope(a),\enc_\ope(m_i)) < \Delta(\enc_\ope(a),\enc_\ope(m_1)) \texttt{..}\\
& \text{\texttt{Group by $\enc_\aes(a)$, Order by $\enc_\ope(a)$}}
\end{align*}
}
Since $\Delta$ is computed over OPE ciphertext, the response from the cloud for $\queryGroupBy$
might not be accurate, as compared to the same query executed over the plaintext data. OPE's
only guarantee is $\enc_\ope(x) < \enc_\ope(y) \leftrightarrow x < y$, hence it does not always follow
that $|\enc_\ope(x) - \enc_\ope(x')| < |\enc_\ope(y) - \enc_\ope(y')| \leftrightarrow |x-x'| < |y-y'|$. In the next
section, we show that this phenomenon occurs frequently, yet the final clusters are very
close to the clusters found using the unencrypted data.

\subsection{Association rule mining (Apriori).}
\begin{algorithm}
\footnotesize
\caption{Apriori association rule mining}
\label{alg:apriori}
\textbf{Input}: \emph{minsup, minconf}, $i$\\
\textbf{Output}: set of rules $\{(X \to Y)\}$\\
\vspace{0.1cm}
$\pk \leftarrow \setup(\kappa)$; $\sk \leftarrow \keygen(\pk)$; $\mak_i \leftarrow
\maskgen(i,\pk)$\\

$L_1 \leftarrow$ \textit{GenerateFrequentItemsetSize1}()\\
$k = 2, B_i = \emptyset$\\
$C_k \leftarrow$ \textit{GenerateCandidates($L_{k-1}$)}\\

\vspace{0.1cm}
\textbf{foreach} $c \in C_k$:\\
$\quad$ $t \leftarrow \queryCount(c)$\\
$\quad$ $B_i = B_i \,\cup\, t$\\

\textbf{foreach} $j \in [1,k]$:\\
$\quad$ $B[j] \leftarrow \computesum(i,\pk,\sk,\mak_i,B_i[j])$\\
$\quad$ extract $c.count$ from $B[j]$\\

\vspace{0.1cm}
$L_k \leftarrow \{ c \in C_k | c.count \geq minsup \}$\\
Increase $k$ and repeat from line 6 until $L_k = \emptyset$\\

\textit{GenerateRules}($\bigcup_k L_k$,\textit{minconf})
\end{algorithm}
An association rule mining algorithm extracts the relationships between attributes that occur
frequently in the data. An association rule has the form $(X \to Y)$ where $X, Y \subseteq A$.
The Apriori algorithm (Algorithm~\ref{alg:apriori}) first determines frequent item-sets
containing a single item using the \textit{GenerateFrequentItemsetSize1} protocol.  The results are merged into larger item-sets
(candidates) using \textit{GenerateCandidates}. The threshold value \emph{minsup} specifies
the lower bound for item-set frequency. These steps are repeated until there is no more
item-set to be found.  Finally, \textit{GenerateRules} generates the outputs by establishing
rules whose confidence values are above $minconf$. The details of
\textit{GenerateFrequentItemsetSize1}, \textit{GenerateCandidates} and \textit{GenerateRules}
can be found in~\cite{apriori}.

\section{Evaluation}
\label{sec:evaluation}
We have implemented the protocols described in the previous sections in order to demonstrate
CloudMine's functionality as well as to preliminarily assess its performance in a hybrid cloud
environment. In particular, the prototype implements the secure sum service and three data
mining algorithms built using this service. It is written in Java, with
cryptographic operations provided by the Crypto++ library~\cite{cryptopp}, OPE and Paillier
encryptions by CryptDB library~\cite{popa11}. Data mining algorithms made use of the
Weka library~\cite{weka}. Communications between data owners and delegates are done via Java
sockets. The source code is available at \url{https://code.google.com/p/cloudmine-sum/}.

\begin{table*}
\footnotesize
\centering
\begin{tabular}{l|l|p{6cm}}
\hline
\textbf{Parameters} & \textbf{Description} & \textbf{Values}\\
\hline \hline
$n$ & number of parties & $2,4,8,16$\\
$k$ & number of delegates & $1,2,4,8,16$\\
$\text{\emph{it}}$ & EC2 instance types & small, medium, large \\
$\text{\emph{bl}}$ & encryption bit length & $512, 1024$\\
$r$ & secure sum request rate & $10,100,200,400,700$\\
$\text{\emph{ds}}$ & dataset & \emph{breast\_cancer, x50\_breast\_cancer, mushroom,
x50\_mushroom, splice,x10\_splice}\\
$\text{alg}$ & data mining algorithm & NaiveBayes, Apriori, K-Mode\\ \hline
\end{tabular}
\caption{List of parameters used in experiments.}
\label{tab:parameters}
\end{table*}

We first experimented with  CloudMine as a stand-alone service. We used \emph{throughput} ---
the number of secure sum operations completed per second measured at the party--- as the
metric.  Next, we evaluated the performance of CloudMine when being used in complex data mining
algorithms. For this, we measured the overall and detailed breakdown of the \emph{running time}
of each data mining algorithm. We ran all experiments on Amazon EC2 platform~\cite{ec2}, using
the parameters as listed in Table~\ref{tab:parameters}. Unless otherwise stated, each delegate
runs on one large EC2 instance, and two parties share one large EC2 instance. In addition,
$n=k=8$ and $\text{\emph{bl}}=1024$. The results presented below are averaged over multiple
runs.

\subsection{Secure Sum Benchmark}
\begin{figure*}
	\hspace*{-1cm}
	\subfloat[Throughput, with varying request rate]{\includegraphics[scale=0.27,angle=-90]{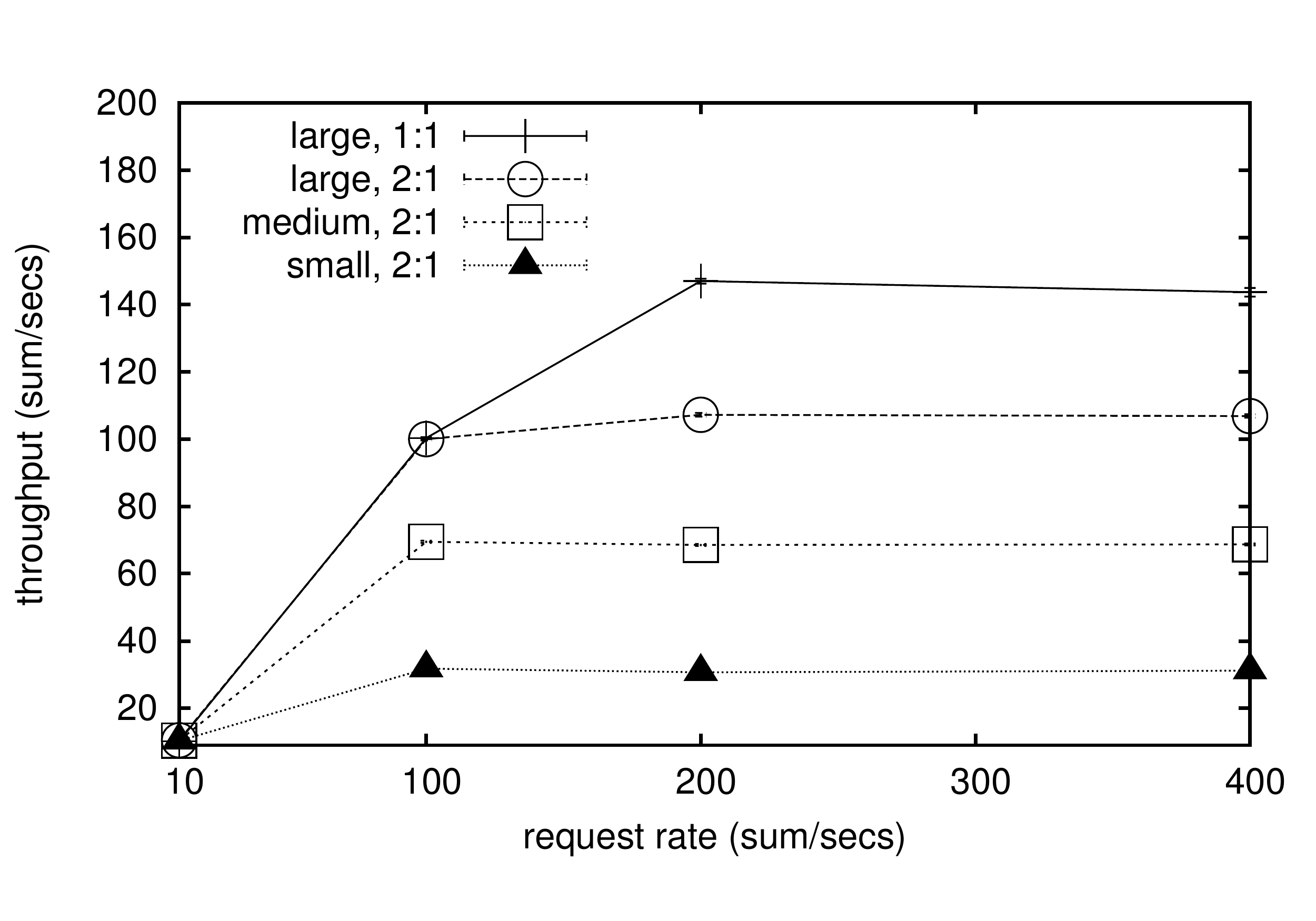}}	
	\subfloat[Throughput at steady state, with varying
$n$]{\includegraphics[scale=0.27,angle=-90]{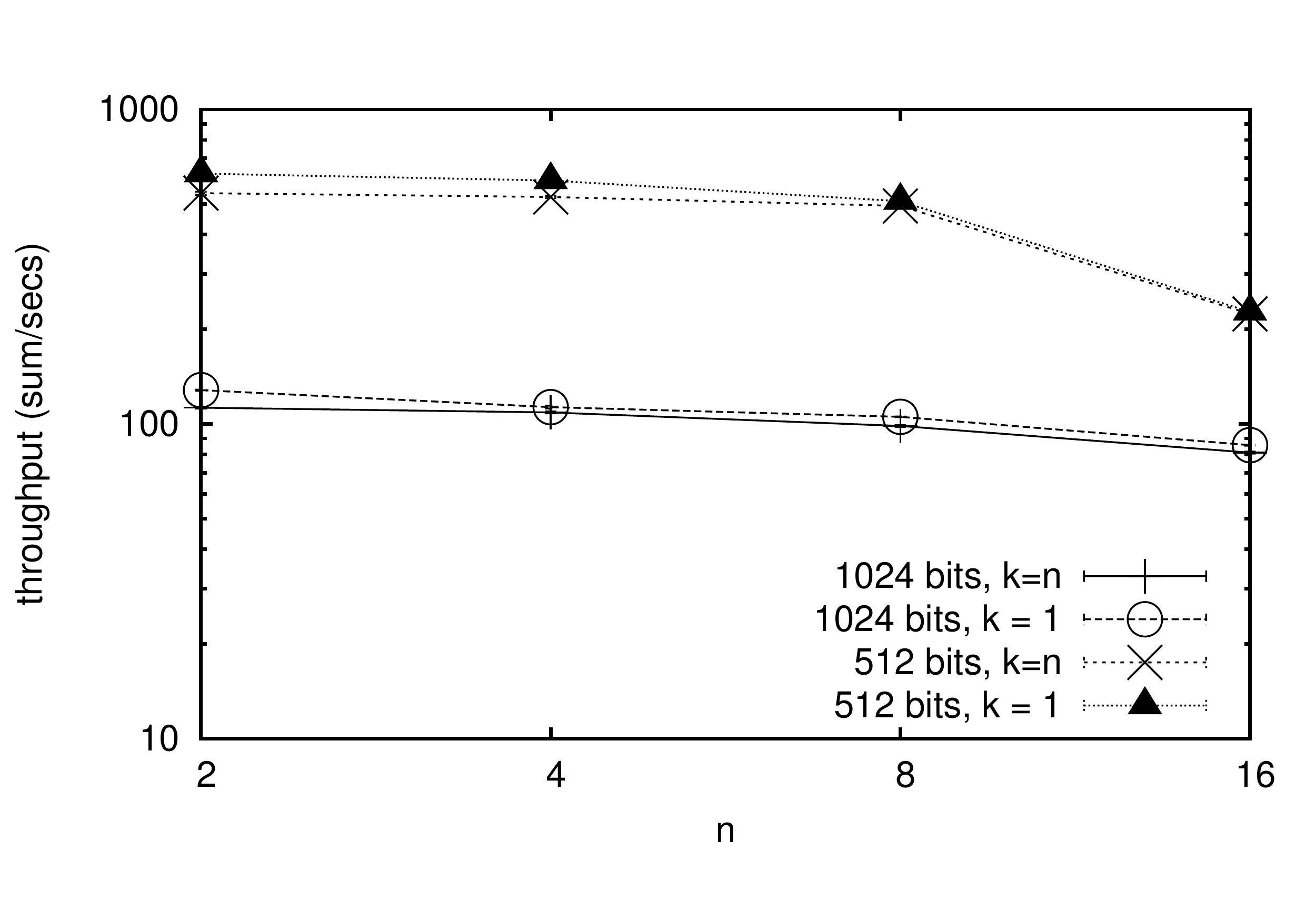}}
	\caption{Secure sum throughput}
	\label{fig:throughput}
\end{figure*}
To benchmark CloudMine service, we varied the frequency at which each party requests for
the service from its delegate. We also varied the types of EC2 instances on which
the party is run, and the number of parties sharing one instance.
Figure~\ref{fig:throughput}[a] shows that throughput reaches its steady state at different values for different
configurations of the party. In particular, the highest throughput is observed at $150$
(sums/sec) when one party occupies one large instance. When two parties share the same
instance, throughput dips to around $110$ (sums/sec). When medium or small instances are
used for the parties, throughput falls even further (the lowest is at $33$ (sums/sec) with
parties sharing small instances). These results indicate with fixed $n$ and $k$, throughput
depends on the computation at the parties, i.e. the more powerful the parties are, the higher
the overall throughput. Furthermore, considering that our prototype implementation has not
been optimized for highly parallel workload, we believe these throughputs are practical for many
real-time applications in which data does not arrive at extremely high rates.

Figure~\ref{fig:throughput}[b] shows how throughput also depends on encryption bit-length
\emph{bl}, the ratio $\frac{k}{n}$ and the number of parties $n$. It can be easily seen that
reducing the encryption bit-length from 1024 to 512 leads to substantial increase in
throughput. This is because Paillier encryption and decryption operations take roughly 1$ms$
when $\text{\emph{bl}}=512$, which rise to $7ms$ with $\text{\emph{bl}}=1024$. The ratio
$\frac{k}{n}$ represents the level of decentralization. When $k=1$, all parties communicate to
one centralized delegate --- the model adopted in~\cite{duan10,shi11,kursawe11}.
When $k=n$, each party has one delegate and each delegate has one party.
The results indicate that throughput is always slightly higher when
$k=1$ than when $k=n$. This means that throughput is mainly determined by the sum
computation, as opposed to be affected by the communication overhead incurred when $k=n$. In
other words, our service supports the decentralization of the multi-party computation with
minimal cost to the overall performance.  Thus, there is no substantial advantage, at least in
terms of throughput, in using a centralized service for secure sum computation; whereas
distributing this private computation over multiple delegates implies the decentralization of
trust, which is a more acceptable model in practice. Finally, as $n$ increases, we can observe a drop
in throughput. This is caused by the computation and communication overhead incurred at the delegates
when $n$ gets larger. We will discuss this overhead in more detail shortly.

\subsection{Data Mining Performance}
\begin{figure*}
\hspace*{-0.5cm}
\subfloat[Encryption and database loading, \emph{alg} =
NaiveBayes]{\includegraphics[scale=0.27,angle=-90]{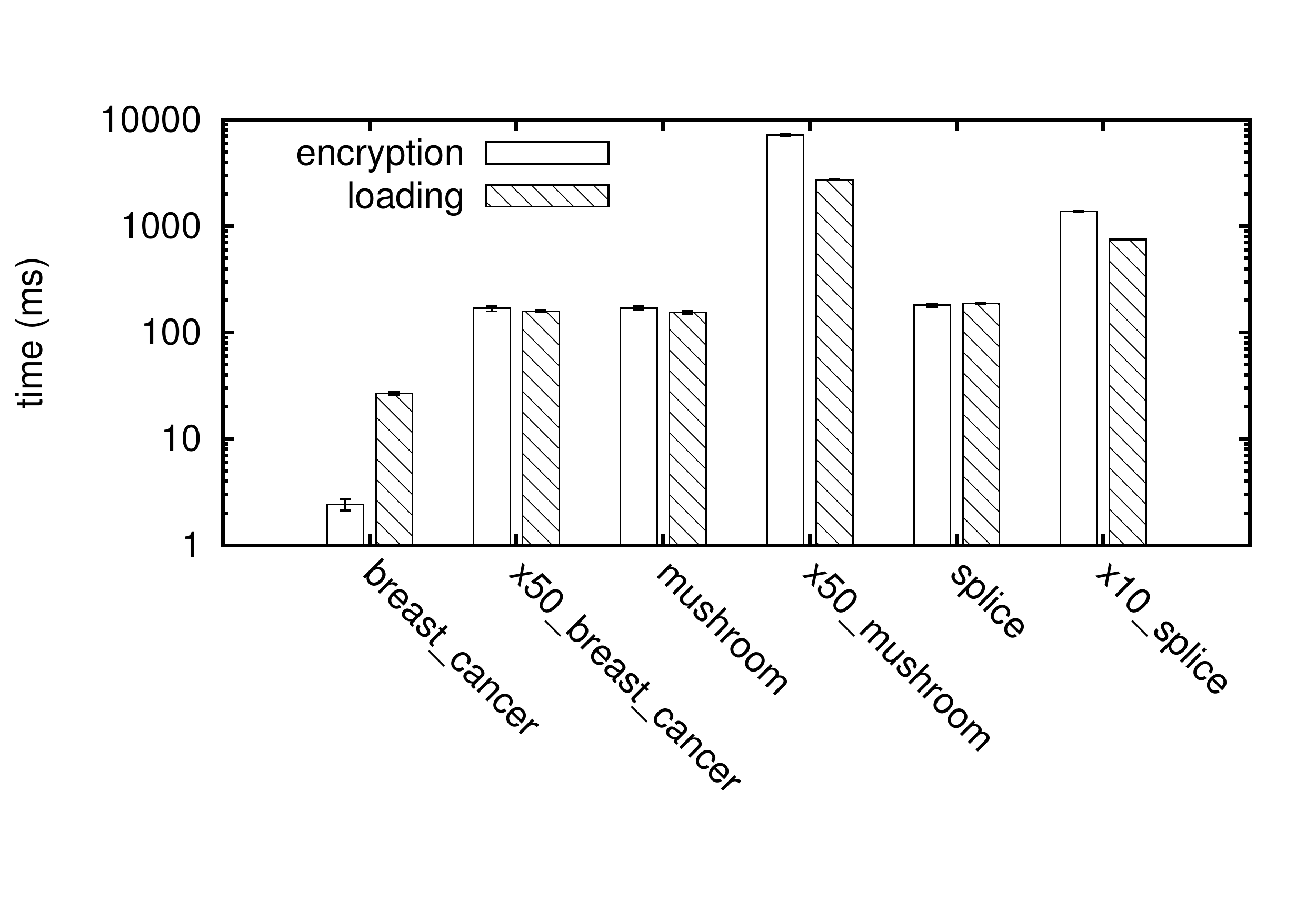}}
\subfloat[Database loading, \emph{alg} = Apriori]{\includegraphics[scale=0.27,angle=-90]{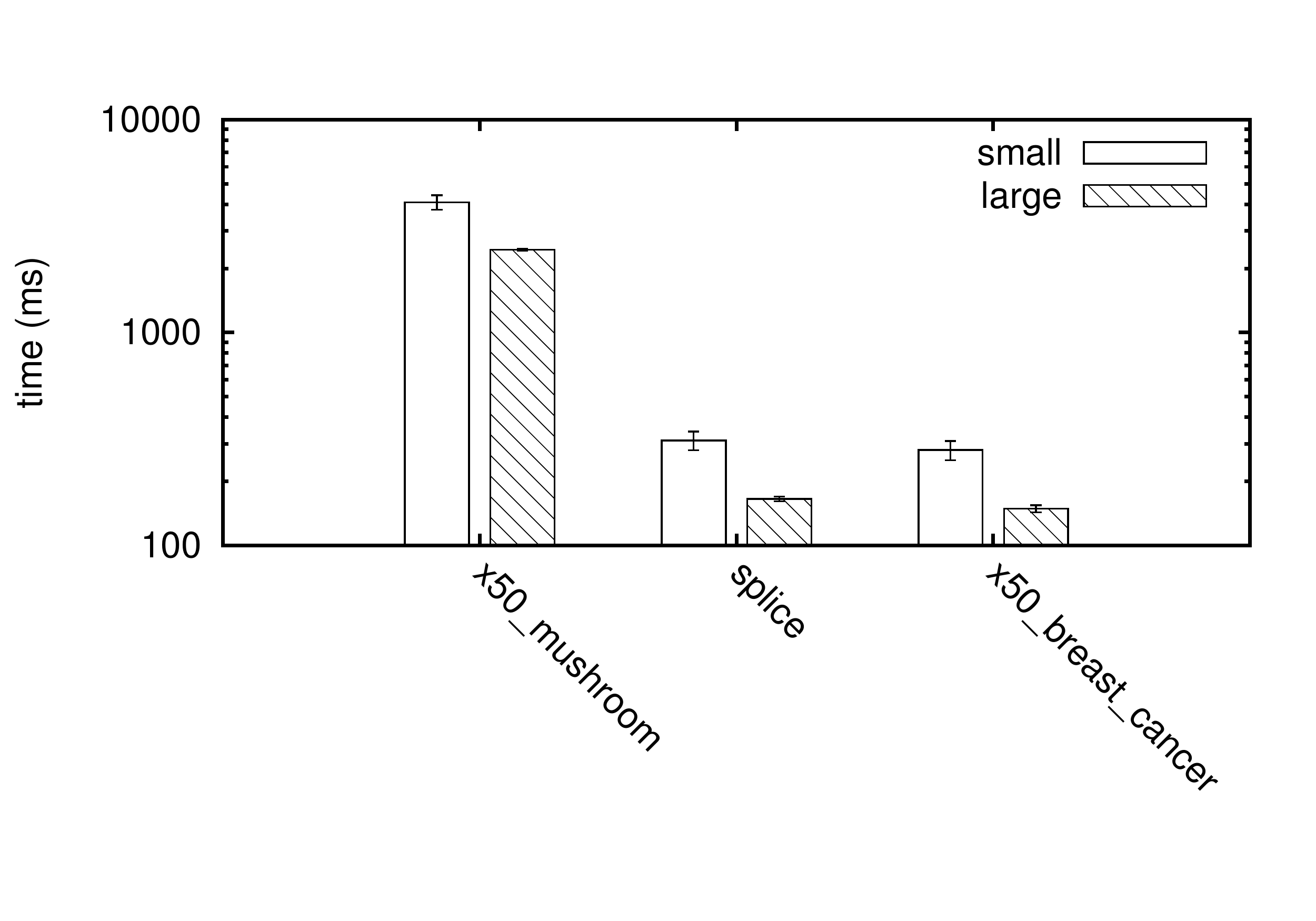}}
\caption{One-time cost}
\label{fig:oneOffCost}
\end{figure*}

We used three standard datasets: \emph{breast\_cancer} (small), \emph{mushroom}
(large, many rows) and \emph{splice} (large, many columns) from~\cite{uciDataset}, and
synthesized larger datasets by extending them with random values from similar distributions.
For instance, \emph{x50\_mushroom} represents the dataset 50-time the size of the original
\emph{mushroom} dataset. The largest dataset consists of $91350$ rows and $23$ columns.

In our prototype, each data owner encrypts its data with AES and OPE and uploads it to the
delegate which then stores it in a MySQL server. We let data owners outsource
all of their data to the cloud, causing larger data query overhead than when parts of
the data are stored locally. The encrypted datasets were as much as $23$
times larger in size than the original, unencrypted ones (for the \emph{x10\_splice} dataset).
We quantify the costs for database encryption at the party and database loading at the
delegate, which incur only once at the beginning, in
terms of the time taken to complete the operations. Figure~\ref{fig:oneOffCost}[a]
illustrates these costs with varying datasets for the NaiveBayes algorithm. It can be seen that
both encryption and loading time are proportional to the data size, and they remain below $8s$
even for the largest dataset. Figure~\ref{fig:oneOffCost}[b] shows the loading time at the delegates
when delegates are running on different types of EC2 instances. Across all datasets, using small
instances results in longer loading time.

\begin{figure}
\centering
{\includegraphics[scale=0.3,angle=-90]{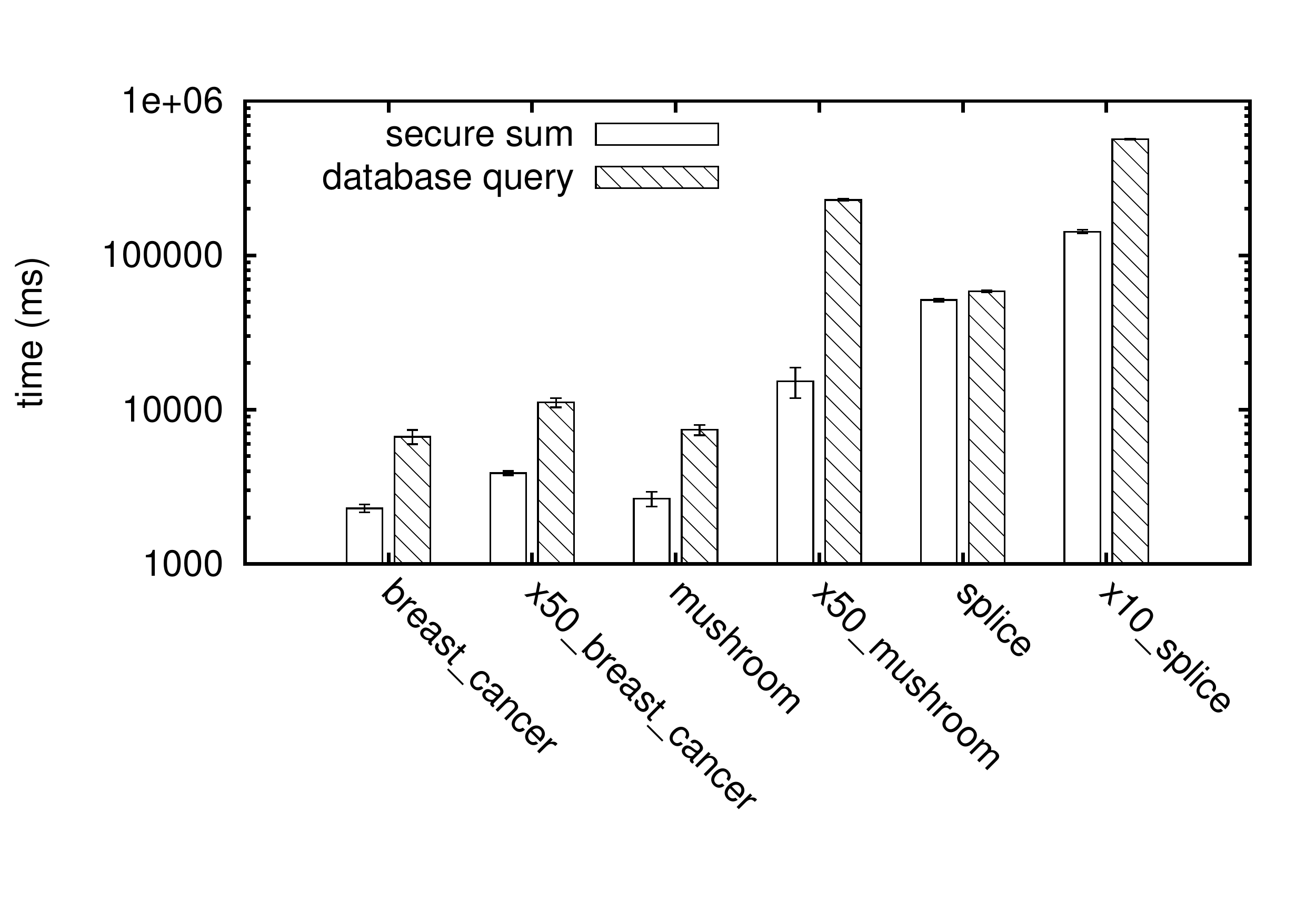}}
\caption{Overall running time, \emph{alg} = Apriori}
\label{fig:overallTime}
\end{figure}
\begin{figure}
\centering
\includegraphics[scale=0.3,angle=-90]{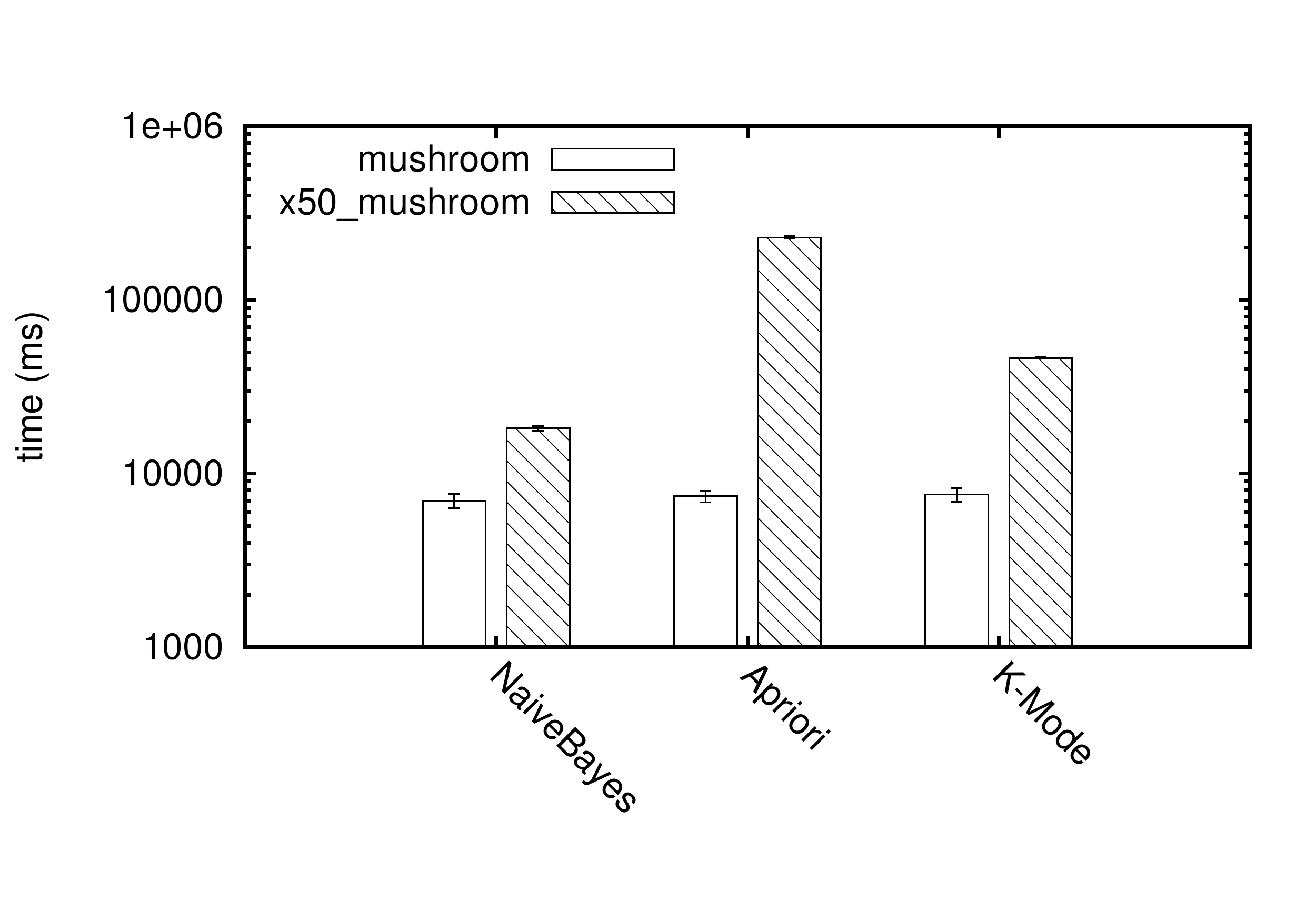}
\caption{Database query time for \emph{ds = mushroom} and \emph{ds = x50\_mushroom datasets}}
\label{fig:queryTime}
\end{figure}

As explained in Section~\ref{sec:dataMining}, a data mining application built using CloudMine
consists of two iterative, interleaving processes: database query and secure sum.
Figure~\ref{fig:overallTime} shows the breakdown costs of these processes --- measured as the
time taken to complete the process --- for Apriori
algorithm. One important observation is that database query time is
always greater than secure sum time. For \emph{x50\_mushroom} dataset, the former takes more
than an order of magnitude longer to complete. The longest experiment (with \emph{x10\_splice}
dataset) took 12 minutes to complete, of which secure sum operation accounted for only 2
minutes. This suggests that when used in real data mining algorithms, the cost of the secure
sum service has small effect on the overall performance. Figure~\ref{fig:queryTime} shows the effect of increasing data size to
the database query time for different algorithms. It can be observed that query time scales
differently for different algorithms. Particularly, Apriori demonstrates the sharpest growth as
compared to NaiveBayes and K-Mode. We attribute this to the intrinsic properties of the data
mining algorithm. Specifically, we observe that in our experiments with Apriori, larger datasets led to more queries being
performed by the delegate (from $132$ with \emph{mushroom} to $4214$ with the \emph{x50\_mushroom} dataset).

\begin{figure}
\centering
\includegraphics[scale=0.3,angle=-90]{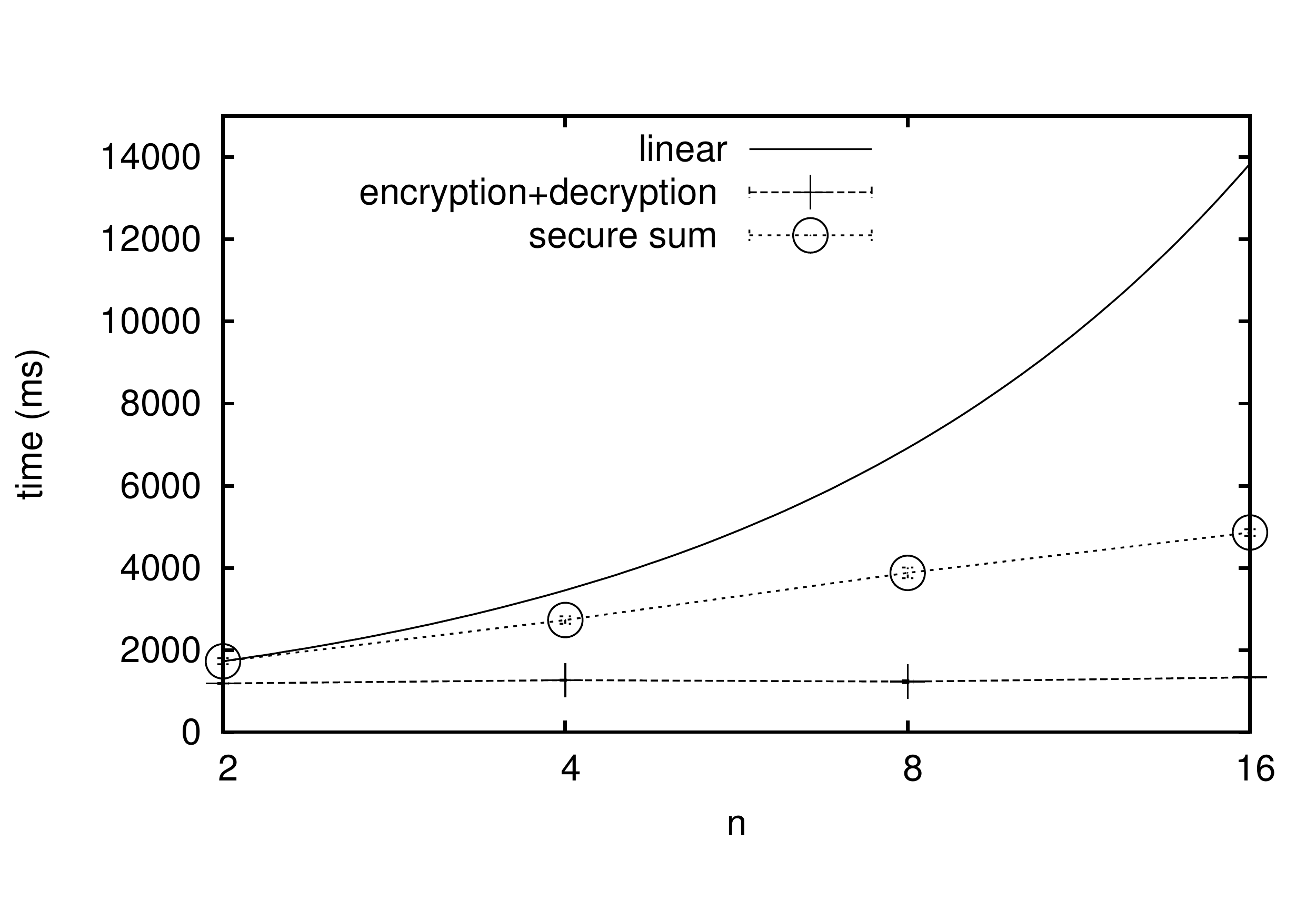}
\caption{Secure sum time for the \emph{x50\_breast\_cancer} dataset. \emph{alg} = Apriori}
\label{fig:smcTimeVaryingParties}
\end{figure}
Finally, we investigated the cost of the secure sum service as being used in data mining
algorithms. Figure~\ref{fig:smcTimeVaryingParties} shows this cost varies with $n$ for the
Apriori algorithm. As the number of parties gets larger, the secure sum cost also
increases, albeit at a sub-linear rate. This is consistent to what has
been observed in Figure~\ref{fig:throughput}[b]. Recall that the cost of a secure sum
operation comprises the encryption/decryption cost at the party and the computation and
communication cost at the delegates. The former is shown in
Figure~\ref{fig:smcTimeVaryingParties} to be almost constant, meaning that the overhead
incurred when $n$ increases can be attributed to the overhead at the delegate. Firsts, each
delegate needs to perform more multiplications when $n$ increases. Second, each will
have to wait longer to receive all the messages from other delegates when $n$ becomes bigger.

\subsubsection{Correctness of K-Mode.} As explained in Section~\ref{sec:dataMining},
the $\queryGroupBy$ protocol in K-Mode may return a different result
as compared to performing the corresponding query locally on the plaintext data. We refer to
this as \emph{mismatched query}, whose error may affect the convergence rate of the algorithm as well as the
final clusters. All of our experiments with K-Mode converged to final modes. To quantify the
differences between clusters found by using CloudMine and what are found using standard K-Mode
over plaintext data, we used an error metric  $\epsilon(C_i, C'_i) = \frac{|\Omega(C_i) -
\Omega(C_i')|}{\Omega(C_i')}$ where $C_i, C'_i$ denote the two clusters and $\Omega(C_i)$ is the mean
squared distance of the members of $C_i$ to the mode. While the average number of mismatched
queries ranges from $0$ (for \emph{mushroom} dataset) to $508.2$ (for \emph{splice} dataset),
the maximum error is $0.03$. This means our protocols yield nearly identical clusters to what
obtained from the standard K-Mode.

\subsection{Discussion.}
The results above have demonstrated that there are overhead incurred by cryptographic operations
when using CloudMine, as compared to when the data owners use their own
infrastructure and directly take part in the multi-party protocol with each other. While these
costs are necessary to provide security in the presence of the delegates, we also remark that
when used in the context of data mining, they become less substantial, and can be more
than offset by the benefits gained from using elastic cloud resources. In particular, let $m$
be the number of secure sum messages sent and received by the data owners during a data mining
algorithm.  Let $\alpha$ be the cryptographic cost for encrypting and decrypting a message
(with additive homomorphic encryption schemes). Let $q$ be the number of database queries and
$c_q$ the CPU cost for each query. The computation overhead at each data owner becomes $O =
(C_d - C) = (\alpha.m
- q.c_q)$ where $C$ is the cost when the data owner uses its own infrastructure. It can be seen
  that $O$ diminishes quickly and becomes negative for larger workloads: more complex data
mining algorithms with high value of $q$ or larger datasets with high $c_q$. It has been shown
in Figure~\ref{fig:overallTime}, for example, that the database query costs may be over an order of
magnitude more than the costs incurred by the secure sum service.





\section{Related Work}
\label{sec:relatedWork}
CloudMine shares common goals with many other works in the area of distributed,
privacy-preserving data analytics. Our work is not based on randomization
approach~\cite{agrawal00} which perturbs the inputs or differential privacy~\cite{dwork06}
approach which adds noise to the outputs. Instead, CloudMine follows the secure multi-party
computation approach~\cite{yao82} in preserving data privacy during computation. It has been
shown that any computation can be done in a private manner, by reducing the computation to a
combination of circuits.  Vaidya et al.~\cite{vaidya03} use generic circuits for evaluating
2-party comparison operation, which is then used for K-Means algorithm over vertically
partitioned data. Yang et al~\cite{yang06} use generic circuits for computing Bayesian networks
on vertically partitioned data. CloudMine does not rely on circuit evaluation, which is either
expensive~\cite{duan10} or is restricted to two-party computation~\cite{huang12}. Instead, it
shares similar model to what is proposed in~\cite{duan10,kursawe11,shi11,rastogi10} which rely
on third-party servers. However, these works focus on specific functions for specific
application domains. In contrast, CloudMine is designed in a service-oriented manner, that can
be flexibly used by a wide range of applications. Furthermore, the adversary model of CloudMine
is stronger than in the aforementioned previous works.

Our delegated computation model is a special case of verifiable computation, in which a client
outsources its computations to a more powerful entity and is able to later verify the outputs.
Theoretical results have shown that any computation can be outsourced with guaranteed input and
output privacy~\cite{gennaro10}. However, a general protocol for outsourced computation is
inefficient~\cite{wang11}. \cite{golle01,goldwasser08} propose to detect cheating and mis-computation
at the expense of data privacy, but they rely on probabilistic
checking and require the client to pre-compute the results or the delegate to commit certain
values. Wang et al.~\cite{wang11,wang11a} propose practical methods to outsource linear programming
to the cloud. However, they consider a single data owner and delegate, as opposed to CloudMine's
multi-party model.

Finally, existing works on security of outsourced databases focus on data
privacy~\cite{popa11}, query freshness~\cite{merkle79,goodrich01} and query
completeness~\cite{li10}. These works complement the protocols we described in
Section~\ref{sec:dataMining} (which deal with data privacy and query completeness).

\section{Conclusions and Future work}
\label{sec:conclusion}
In this paper, we have described a cloud-based service, named CloudMine, which allows multiple
data owners to carry out analytics over their joint data in a privacy-preserving manner. The
computation is outsourced to a number of independent clouds (or delegates). CloudMine protects
data privacy and ensures correctness of the computation against the standard semi-honest model
of the data owners, and against the curious-and-lazy delegate model. CloudMine supports three
analytic functions: secure sum, secure set union and intersection, and secure scalar product.
These primitives can be used to implement a wide range of complex data mining algorithms. We
demonstrated this by showing how a simple instance of CloudMine (the secure sum service) can be
used in a hybrid cloud environment for the classification, association rule mining and
clustering algorithms. We discussed the mechanisms designed to ensure privacy when the data is
stored in a public cloud. Finally, we implemented a prototype of CloudMine's secure sum service
and evaluated the performance of the service as a stand-alone application and as part of
complex data mining applications. The results demonstrate the service's practical performance,
and show that it provides privacy with little cost to the overall performance for workloads that are
inherently computationally intensive.

Our current prototype has not implemented the protocols for bootstrapping the CloudMine
service. Dynamic group membership may affect the service and its applications in interesting
ways.  Incorporating and evaluating these protocols, and optimization of the overall
implementation are parts of our immediate plan for future work. We also plan to implement the
protocols for secure set and scalar product services. For the former, particularly, we intend to
investigate how existing protocols for private set intersection (which scale better than our
current protocol) can be modified to work in our delegate settings. Once being
equipped with these higher-level primitives, we can start looking at more complex applications
such as collaborative filtering. Additionally, we plan to explore if the automated scaling
features offered by some cloud platforms could improve the performance of the service,
especially under intensive workloads. Finally, we would like to incorporate differential
privacy techniques into the service and investigate the maximum privacy budget needed to
realize any given data mining algorithm.

\bibliographystyle{plain}
{
\bibliography{paper}
}
\end{document}